\documentstyle[12pt]{article}

\def\beq{\begin{equation}}
\def\eeq{\end{equation}}

\catcode`\@=11 % This allows us to modify PLAIN macros.

\def\lsim{\mathrel{\mathpalette\@versim<}}
\def\gsim{\mathrel{\mathpalette\@versim>}}
\def\@versim#1#2{\vcenter{\offinterlineskip
    \ialign{$\m@th#1\hfil##\hfil$\crcr#2\crcr\sim\crcr } }}

\def\t1{{\tilde 1}}

\def\GeV{\,{\rm GeV}}
\def\TeV{\,{\rm TeV}}

\def\NPB#1#2#3{Nucl. Phys. B {\bf#1} (19#2) #3}
\def\PLB#1#2#3{Phys. Lett. B {\bf#1} (19#2) #3}

\def\PRD#1#2#3{Phys. Rev. D {\bf#1} (19#2) #3}
\def\PRL#1#2#3{Phys. Rev. Lett. {\bf#1} (19#2) #3}

\def\MODA#1#2#3{Mod. Phys. Lett. A {\bf#1} (19#2) #3}

\begin{document}
% TH format
\begin{flushright}
\baselineskip=12pt
%{CERN-TH.????/94}\\
%{CTP-TAMU-52/94}\\
%{ACT-15/94}\\
{MIU-THP-71/95}\\
\end{flushright}

\begin{center}
\vglue 1 cm
{\Huge\bf Upper bounds on supersymmetry breaking from
gauge coupling unification\\}
\vglue 1 cm
{S. KELLEY}
\vglue 0.2cm
{\em Department of Physics, Maharishi International University\\
Fairfield, Iowa~52557--1069\\}
\baselineskip=12pt

\vglue 1 cm
ABSTRACT
\end{center}
%\vglue 0.5cm
{\rightskip=3pc
 \leftskip=3pc
\noindent
\baselineskip=20pt
I derive conservative upper bounds on the supersymmetry
breaking parameter $m_{1/2}$ as a function of the strong
coupling in the Standard Supersymmetric Model (SSM) using
gauge coupling unification.  I find that over more than
$99\%$ of the parameter space, $\alpha_3>0.120$ implies
that $m_{1/2}$ is below $10\TeV$ and $\alpha_3>0.129$ implies
that $m_{1/2}$ is below $1\TeV$.  I express the variation of
these bounds over the SSM parameter space with a numerical
coefficient, $c$.  I also find that in the SSM, a reasonable
value of $50\GeV<m_{1/2}<1\TeV$ requires $\alpha_3>0.119$
over the whole parameter space.  These bounds are particularly
sensitive to the value of $\sin^2\theta_W=0.2317\pm0.0005$
used in the calculation.  In more realistic models, heavy
thresholds and gravitational effects will modify this result.
Although these effects are theoretically calculable in specific
models, more realistic models contain many unknown parameters
in practice.  I illustrate this point with minimal
supersymmetric $SU(5)$ where the combined constraints of gauge
coupling unification and proton decay require $\alpha_3>0.119$
for $m_{1/2}<1\TeV$ and the upper bound on the supersymmetry
breaking scale is greatly relaxed.
}

\vspace{1cm}

%\vspace{0.5cm}
% TH format
\begin{flushleft}
\baselineskip=12pt
%{CERN-TH.????/94}\\
%{CTP-TAMU-52/94}\\
%{ACT-15/94}\\
{MIU-THP-71/95}\\
February 1995
\end{flushleft}

\vfill\eject
\setcounter{page}{1}
\pagestyle{plain}
\baselineskip=14pt

Gauge coupling unification \cite{georgi} applied to precision LEP
measurements has provided strong motivation for supersymmetry,
\cite{akl}
and inspired numerous attempts to extract the supersymmetry breaking scale and
constrain the parameter space of specific models \cite{deboer}.  However,
definite conclusions only result from very specific models:
threshold effects and gravitational corrections make model-independent
statements difficult \cite{hall}.
In this paper, I find interesting constraints from gauge coupling
unification within the specific framework of the Standard Supersymmetric
Model (SSM) as defined in reference \cite{SSM} and the minimal
supersymmetric $SU(5)$ model.  Although these models are an excellent
beginning, they can hardly be considered ultimate theories as neither
include gravity and the fine-tuning problem in minimal supersymmetric
$SU(5)$ requires some modification of the GUT structure.
However, there is hope that similar
constraints could be derived for specific realistic string models
where the additional threshold and gravitational effects are in
principle calculable.

Direct searches for supersymmetric particles have continually
increased the lower bounds on their masses.  However, the
only upper bounds on supersymmetric masses come from
naturalness arguments \cite{nat} or cosmological constraints on
the LSP relic density \cite{lsp}.  Although compelling, the naturalness
bounds are not rigorous, and whether the upper bound on
supersymmetric masses is $1\TeV$, $10\TeV$ or even $100\TeV$ is
not clear and somewhat a matter of taste. The cosmological bounds
can be evaded, for example by breaking R-parity.  It would
be extremely useful to have some other method of bounding the
supersymmetric masses.
In this paper, I focus on a first step in this direction
by deriving
an upper bound on the soft supersymmetry breaking parameter
$m_{1/2}$ as a function of the strong coupling in the Standard
Supersymmetric Model (SSM).
Our approach attempts
a general and analytic analysis to reveal the underlying
physics, and is meant to complement the many numerical searches of
the SSM parameter space in the literature \cite{deboer}.
Unless otherwise indicated,
our notation assumes all gauge couplings are renormalized at $m_Z$
in the $\overline{MS}$ scheme.

Ignoring gravitational effects, gauge coupling unification in the SSM
gives a simple prediction for the soft supersymmetry breaking
parameter $m_{1/2}$.  Restricting attention to values of
$m_{1/2}$ for which all the supersymmetric thresholds are
above $m_Z$, this prediction is \cite{detail}:
\begin{equation}
\ln({{m_{1/2}}\over{m_Z}})=-X
+{{7\pi}\over{\alpha_3}}
+7\ln\left[{{\alpha_3(m_{\tilde g})}\over{\alpha_2(m_{\tilde w})}}\right]
-\ln(c_{\tilde w})
\end{equation}
where the gaugino masses are given by
\begin{equation}
m_{\tilde i}=c_{\tilde i}~m_{1/2}
={{\alpha_i(m_{\tilde i})}\over{\alpha_G}}m_{1/2}~,
\end{equation}
$\alpha_G$ is the unified coupling at the unification
scale,
and $f(y,w)$ includes the threshold effects of squarks and sleptons
under simple assumptions for the form of the stop mass matrix \cite{detail}
\begin{eqnarray}
f(y,w)&=&{{15}\over 8}\ln(\sqrt{c_{\tilde q}+y})
-{9\over 4}\ln(\sqrt{c_{\tilde e_l}+y})
+{3\over 2}\ln(\sqrt{c_{\tilde e_r}+y})\nonumber\\
&&-{{19}\over{48}}\ln(\sqrt{c_{\tilde q}+y+w})
-{{35}\over{48}}\ln(\sqrt{c_{\tilde q}+y-w})\ ,
\end{eqnarray}
where $y\equiv m^2_0/m_{1/2}^2$ and $w\equiv\bar m^2/m^2_{1/2}$
with $\bar m^2$ representing the
off-diagonal elements in the stop mass matrix.
The quantity $X$ is given by
\begin{eqnarray}
X&=&-{{15\pi}\over{\alpha_{em}}}
\left[.2+\delta_s(gauge)+\delta_s(Yukawa)-\sin^2\theta_W\right]\nonumber\\
&&-f(y,w)
+{9\over 4}\ln({{m_t}\over{m_Z}})+3\ln({{\mu}\over{m_Z}})
+{3\over4}\ln({{m_H}\over{m_Z}})~,
\end{eqnarray}
where $\mu$ is the dimensionful higgs mixing term in the
SSM superpotential and $m_H$ represents the mass of the
charged, pseudoscalar and heavy scalar higgs which are
nearly degenerate in the SSM.
The gauge numeric corrections to two loop accuracy are \cite{ver}
\begin{equation}
\delta_s(gauge)=0.00127+0.01480\alpha_3~,
\end{equation}
and numerical calculations give $-0.0004<\delta_s(Yukawa)<0$ \cite{ldelta}.
To derive the most conservative upper bound on $m_{1/2}$, I
will want to maximize the right side of Equation (1), and
therefore minimize $X$.  Note, by definition, there are no
GUT or intermediate thresholds in the SSM \cite{SSM}.

The calculation is most sensitive to $\sin^2\theta_W$,
for which the central value has come down in the last year to \cite{lsin}
\begin{equation}
\sin^2\theta_W=0.2317\pm 0.0005~.
\label{eq:5}
\end{equation}
Bounds on other parameters I will use are:
\begin{equation}
\alpha_{em}={1\over{127.9\pm.0.1}}\quad\quad
131\GeV<m_t<190\GeV~.
\label{eq:6}
\end{equation}

Although the most conservative choice for $\mu$ and $m_H$
would be to take them below $m_Z$, values of these
parameters in the SSM turn out to be nearly proportional
to $m_{1/2}$ because of the correlations introduced by radiative
electroweak symmetry breaking.  In fact, $\mu$ and $m_H$ are
determined as a function of the five parameters of the SSM:
$m_t,\tan\beta,m_{1/2},\xi_0,\xi_A$.  The explicit
form of the tree level expressions for $\mu$ and $m_H$ show that
they both approximately scale with $m_{1/2}$:
\begin{equation}
m^2_{\mu,H}=a_{\mu,H}m_{1/2}^2+b_{\mu,H}m_W^2~.
\end{equation}
I use a parameter, $c$, to encode the dependence
on $\mu$ and $m_H$ in a simple way by the equation
\begin{equation}
{15\over4}\ln({{c~m_{1/2}}\over{m_Z}})\equiv 3\ln({{\mu}\over{m_Z}})
+{3\over4}\ln({{m_H}\over{m_Z}})~.
\end{equation}
Note that throughout this paper, the logs associated with a
particular threshold correction are understood to be zero if
the threshold is below $m_Z$.
A Monte Carlo search of parameter space reveals that $c>0.5$
for more than $99\%$ of the points considered.
Our search uses $m_{1/2}=1\TeV$ and
randomly selects $10,000$ points for each sign of $\mu$
over the four dimensional parameter space defined by
$130\GeV<m_t<190\GeV$, $1<\tan\beta<50$, $-10<\xi_0<10$, and
$-2|\xi_0|<\xi_A<2|\xi_0|$.  The parameters $m_t,\xi_0,\xi_A$ are
searched on a linear scale while the parameter $\tan\beta$ is
searched on a logarithmic scale.

To reliably extract $m_{1/2}$, the dependence of the gaugino masses
on $m_{1/2}$ (the EGM effect) must be carefully taken into account \cite{zich}.
To derive the most conservative upper bound on $m_{1/2}$, I
must use an upper bound on $\alpha_3(m_{\tilde g})$, and
and a lower bound on $\alpha_2(m_{\tilde w})$.
The simplest way to extract the gauge couplings renormalized
at the gaugino mass is to iterate the expression
\begin{equation}
\alpha_i(m_{\tilde i})=
{{\alpha_i}\over{1-\alpha_i{{b_i}\over{2\pi}}
\ln({{\alpha_i(m_{\tilde i})m_{1/2}}\over{\alpha_Gm_Z}})}}
\end{equation}
along with the expression for the gaugino mass, Equation (2), to a solution.
Since these values for the gauge couplings increase with $b_i$,
I take the maximum value of $b_3$ below the gluino threshold
and the minimum value of $b_2$ below the wino threshold to
derive the most conservative bounds.
Since in the SSM, the squarks cannot be much lighter than
the gluino, I use $b_3=-7$.  Below the wino threshold,
I use the minimum $b_2=-19/6$.
Numerical calculations
show that an upper bound on $\alpha_G$ gives the most conservative
bound for $m_{1/2}$.
I obtain this upper bound on $\alpha_G$ by extrapolating
the hypercharge coupling $\alpha_y(m_Z)$ to the scale $10^{17}\GeV$ using the
maximum
$b_y=33/5$.  This results in $\alpha_G<0.0454$ after numerically
correcting for two-loop effects.

Finding an upper bound on $f(y,w)$ depends on the physically
acceptable range of the
variables $y$ and $w$.  Although the minimum value of
$f(y,w)=-0.025$ has already been determined \cite{constrain}, the maximum
value is not as clear cut.
For fixed $w$, $f(y,w)$ approaches zero as $y$ becomes
very large.  Maximization with respect to values of
$w$ yielding positive $m_{\tilde t}^2$ gives a
maximum of $f(y,w)=35ln(\sqrt{y})/48$
at $w=c_{\tilde q}+y$ as $y$
becomes very large.  However, a more physically reasonable
choice is $|w|<dm_t/m_{1/2}$ where $d$ is a positive
numeric coefficient of order one.  Using this form for $w$,
I have numerically scanned
over values of $d<10$ to determine that, for $m_{1/2}>1\TeV$,
$f(y,w)<0.5$, which I will use in the calculation.  It is
the fact that the $SU(3),SU(2),U(1)$ beta functions are all equal
for an entire generation coupled with the
assumption of universal soft supersymmetry breaking which severely
restricts the range of $f(y,w)$ and makes the results of gauge
coupling unification approximately independent of the universal
scalar mass $m_0$.

Putting all this together gives equations which can be
iterated to find an upper bound on $m_{1/2}$ as a function
of the strong coupling.
The choice of parameters I make to minimize $X$ and thereby
obtain a conservative bound are:
\begin{equation}
\alpha_{em}={1\over{127.8}}\quad m_t=131\GeV\quad f(y,w)=0.5\quad
\alpha_G=0.0454~.
\end{equation}
The bound using $c=0.5$ is plotted as solid lines
in Figure 1
for the central and $1-\sigma$ values of $\sin^2\theta_W$.
If $\alpha_3>0.120$, $m_{1/2}$ is constrained to be less
than $10\TeV$ and if $\alpha_3>0.129$, $m_{1/2}$ is constrained
to be less than $1\TeV$.  These solid lines would move downward
as $c$ increases giving tighter bounds on $m_{1/2}$ and would
move upward as $c$ decreases giving looser bounds on $m_{1/2}$.
To quantify this effect, in Figure 2 I indicate the value of $\alpha_3$,
as a function of $c$, above which $m_{1/2}$ is bounded by $1\TeV,10\TeV$
with solid,dashed lines.  In our Monte Carlo, among the
12,231 out of 20,000 points which have perturbative Yukawas and
a stable electoweak breaking minimum, the values of $c$ are
distributed as follows: $62.9\%$ give $c>2$, $29.9\%$ give
$2>c>1$, $6.4\%$ give $1>c>0.5$, $0.7\%$ give $0.5>c>0.2$, and
$0.1\%$ give $c<0.2$.  The points with $c<0.5$ all have
$\mu/m_{1/2}<0.4$ but rarely have $m_H/m_{1/2}<1$.  This observation
and the form of Equation (9) indicate that the regions with
small $c$ are the regions where $\mu/m_{1/2}$ is fine-tuned
to be small.

Previous work pursued an alternative to finding an upper
bound on the supersymmetry breaking scale and derived a
lower bound on $\alpha_3$ as a function of $m_{1/2}$ \cite{constrain}.
For values of $m_{1/2}$ giving wino masses above
$m_Z$ this derivation followed from Equation (1).  For
values of $m_{1/2}$ giving wino masses below $m_Z$,
the relation becomes
\begin{equation}
\ln({{m_{1/2}}\over{m_Z}})={1\over7}X
-{{\pi}\over{\alpha_3}}
-\ln(c_{\tilde g})~.
\end{equation}
However, for both cases, the most conservative bound on
$\alpha_3$ results from maximizing $X$, minimizing $c_{\tilde g}$ and
maximizing $c_{\tilde w}$.
This is accomplished by taking $b_3=-7$, $b_2=-1/3$, and
\begin{equation}
\alpha_{em}={1\over{128.0}}\quad\quad m_t=190\GeV\quad\quad
f(y,w)=-0.025\quad\quad
\mu=m_H=1\TeV~.
\end{equation}
I use a minimum value of $\alpha_G$ in Equation (1) and a
maximum value of $\alpha_G$ in Equation (12) from the range
\begin{equation}
{3\over{20\alpha_{em}}}+{3\over{5\alpha_3}}-0.7<
{1\over{\alpha_G}}<
{3\over{20\alpha_{em}}}+{3\over{5\alpha_3}}+1.4
\end{equation}
obtained from gauge coupling unification
with a $3\TeV$ bound on the supersymmetric thresholds.

The resulting
bound on $\alpha_3$ is shown for the central and $1-\sigma$ values
of $\sin^2\theta_W$ by wavy lines in Figure 1.
To have $50\GeV<m_{1/2}<1\TeV$ requires $\alpha_3>0.119$.
Note the differences in the approaches leading
to the upper bound on $m_{1/2}$ indicated by solid lines
and the lower bound on $\alpha_3$ indicated by wavy lines.
The upper bound on $m_{1/2}$ is obtained by maximizing the
right side of Equation (1) with no bounds on
the supersymmetric thresholds.  The lower bound on
$\alpha_3$ is obtained by minimizing the right side of
Equation (1) and maximizing the right side of Equation (12) assuming that
$m_{1/2},\mu,m_H<1\TeV$.
The ultimate reason for this
difference is that the supersymmetric threshold corrections
to $\sin^2\theta_W$ in the SSM are bounded from above by the assumption
of universal soft supersymmetry breaking but are only bounded
from below by naturalness.

As a simple example of modifications required in more realistic
models, consider minimal supersymmetric $SU(5)$.
Although the predictions of gauge coupling unification
depend on the unknown superheavy scales, one can extract
the mass of the superheavy proton decay mediating
triplets as a function of the low-energy couplings \cite{ENR,mur,ver}.
Unfortunately, this leaves no prediction for the supersymmetry
breaking scale.
However, gauge coupling constant unification, a $1\TeV$ naturalness
bound, and limits on the proton lifetime can be combined to
derive a lower bound on $\alpha_3$ \cite{ver}.  Using the
value of $\sin^2\theta_W$ in Equation (6)
gives a bound of $\alpha_3>0.119$ in minimal supersymmteric
$SU(5)$.  Careful study \cite{kras} reveals that in minimal
supersymmetric $SU(5)$, the supersymmetry breaking scale could
be as large as $10^8\GeV$ and extensions of the minimal model
further relax this bound.

In addition to the theoretical
prejudice for a naturalness bound on supersymmetry, this
calculation offers the possibility of using coupling constant
unification to place a bound on $m_{1/2}$ in specific models.
The excellent agreement of gauge coupling unification with
the SSM, combined with the sharpening and shifting measurements
of the low-energy couplings provide interesting speculation.
On the one hand, apart from the highly unlikely possibility of a
light gluino \cite{lightg},
measurements of the strong coupling from
deep-inelastic scattering \cite{deep}, charmonium \cite{charm},
and $\Upsilon$ \cite{bottom}
give low values of $\alpha_3$ which in the SSM and minimal supersymmetric
$SU(5)$ require unnaturally high values of $m_{1/2}$.
On the other hand, the values of $\alpha_3$ from jet shapes
at LEP are increasing to values which ensure a low value of
$m_{1/2}$ in the SSM.  To complicate matters, the
value of $\sin^2\theta_W$ from SLD \cite{SLD} is lower than that from
LEP introducing more uncertainty in this critical input to
gauge coupling unification calculations.

Even though these types of conclusions can so far only be reached
in very simple models like the SSM and minimal supersymmetric
$SU(5)$, the dependence of the results on the unfolding experimental
situation holds for some the excitement of a close horse race.
The observation that low-energy limits of string models
come very close to the simple models considered in this paper,
and the ability of string theory to quantify gravitational
corrections to the predictions of gauge coupling unification,
offers hope of testing the physics of unified theories using
precision low-energy measurements.

\section*{Acknowledgments}
I thank J. Lopez, A. Petermann, K. Yuan, and V. Ziegler for their insights.
I especially thank A. Zichichi and D. Nanopoulos for suggesting this project
and their support.

\newpage
\section*{Figure Captions}

%begin{figure}[p]
%\vspace{6in}
%\special{psfile=nfig1.eps hscale=100 vscale=100 hoffset=-50 voffset=-100}
%\special{postscript\input nfig1.eps}
%\caption{
\medskip
Figure 1:  Solid lines indicate the
upper bound on the supersymmetry breaking scale from gauge coupling
unification in the SSM for central
and $1-\sigma$ values of $sin^2\theta=0.2317\pm0.0005$ with c=0.5.
The supersymmetry breaking parameter $m_{1/2}$ is below
$10\TeV$ for $\alpha_3>0.120$ and is below $1\TeV$ for
$\alpha_3>0.129$.
Wavy lines indicate a lower bound on the strong coupling
from gauge coupling unification assuming a naturalness bound
on the scale of superysmmetry breaking.  Reasonable values
of $50\GeV<m_{1/2}<1\TeV$ require $\alpha_3>0.119$.
%}
%\label{Fig1}
%\end{figure}
%\clearpage

%\newpage

%\begin{figure}[p]
%\vspace{6in}
%\special{psfile=nfig2.eps hscale=100 vscale=100 hoffset=-50 voffset=-100}
%\caption{
\bigskip
\noindent
Figure 2:  The value of $\alpha_3$, as a function of $c$, above which $m_{1/2}$
is bounded by $1\TeV,10\TeV$ is indicated by solid,dashed lines
in the SSM for central
and $1-\sigma$ values of $sin^2\theta=0.2317\pm0.0005$.  The
most conservative choice, $\mu,m_H<m_Z$ corresponds to $c=0$.  A
Monte Carlo search of the SSM parameter space finds that over $99\%$ of
the points considered give $c>0.5$.
%}
%\label{Fig2}
%\end{figure}
%\clearpage


\newpage
\begin{thebibliography}{99}
\bibitem{georgi} H. Georgi, H. Quinn and S. Weinberg, \PRL{33}{74}{451}.
\bibitem{akl} J. Ellis, S. Kelley and D. V. Nanopoulos, \PLB{249}{90}{441};
P. Langacker and M. Luo, \PRD{44}{91}{817};
U. Amaldi, W. de Boer and H. Furstenau, \PLB{260}{91}{447};
F. Anselmo, L. Cifarelli, A. Peterman and A. Zichichi, Nouvo Cimento
A {\bf 104} (1991) 1817.
\bibitem{deboer} W. de Boer, R. Ehret, W Oberschulte and D. I. Kazakov,
IEKP-KA/94-05, hep-ph 9405342 and references therein.
\bibitem{hall} L. J. Hall and U. Sarid, LBL preprint LBL-32905 (1992).
\bibitem{SSM} S. Kelley, J. L. Lopez, D. V. Nanopoulos, H. Pois and K. Yuan,
\NPB{398}{93}{3};
S. Kelley, MIU preprint MIU-THP-93/65, hep-ph 9310218.
\bibitem{nat} J. Ellis, K. Enqvist, D. V. Nanopoulos, and F. Zwirner
\MODA{1}{86}{57};
R. Barbieri and G. F. Giudice, \NPB{306}{88}{63}.
\bibitem{lsp} See for example: J. Ellis and F. Zwirner, \NPB{338}{90}{317};
S. Kelley, J. L. Lopez, D. V. Nanopoulos, H. Pois and K. Yuan,
\PRD{47}{93}{2461}.
\bibitem{detail} J. Ellis, S. Kelley and D. V. Nanopoulos,
\NPB{373}{92}{55}.
\bibitem{ver} J. S. Hagelin, S. Kelley, and V. Ziegler,
MIU preprint MIU-THP-94/68, hep-ph 9406366.
\bibitem{ldelta} P. Langacker and N. Polonsky, \PRD{47}{93}{4028}.
\bibitem{lsin} P. Langacker, UPR-0624T, hep-ph 9408310.
\bibitem{zich} F. Anselmo, L. Cifarelli, A. Petermann and A. Zichichi,
Nouvo Cimento A {\bf 105} (1992) 581.
\bibitem{constrain} J. Ellis, S. Kelley and D. V. Nanopoulos,
\PLB{287}{92}{95}.
\bibitem{ENR} J. Ellis, D. V. Nanopoulos and S. Rudaz, \NPB{202}{82}{43}.
\bibitem{mur} J. Hisano, H. Murayama and J. Yanagida, \NPB{402}{93}{46}.
\bibitem{kras} N. V. Krasnikov, hep-th 9501205.
\bibitem{lightg} I. Antoniadis, J. Ellis and D. V. Nanopoulos,
\PLB{262}{91}{109};
L. Clavelli, P. W. Coulter and K. Yuan, \PRD{47}{93}{1973};
J. L. Lopez, D. V. Nanopoulos and X. Wang, \PLB{313}{93}{241}.
\bibitem{deep} J. Blumlein and J. Botts, \PLB{325}{94}{190}.
\bibitem{charm} A. X. Khadra, G. Hockney, A. S. Kronfeld and P. B.
Mackenzie, \PRL{69}{92}{729}.
\bibitem{bottom} C. T. H. Davies, K. Hornbostel, G. P. Lepage,
A. Lidsey, J. Shigemitsu and J. Sloan, OHSTPY-HEP-T-94-013,
FSU-SCRI-94-79, hep-ph 9408328.
\bibitem{SLD} SLD: K. Abe {\it et al}., SLAC-PUB-6456.
\end{thebibliography}
\end{document}